\documentstyle[10pt, aas2pp4]{article}

\newcommand\nion[2]{#1\,\lowercase{{\sc #2}}}
\newcommand\wave[1]{\mbox{$\lambda$#1\,\AA}}
\def\kmsec{\mbox{km~s$^{\rm -1}$}}
\def\teff{\mbox{T$_{\rm eff}$}}
\def\BmV0{\mbox{$(B-V)^{\rm 0}$}}
\def\VmK0{\mbox{$(V-K)^{\rm 0}$}}
\def\MV0{\mbox{$M_{\rm V}^{\rm 0}$}}

\def\msun{M$_{\odot}$}

\def\etal{\mbox{{\it et al.}}}
\def\eg{\mbox{{\it e.g.}}}

\def\nhi{\noindent \hangindent=0.5truein} 
\def\farcs{\hbox{$\> .\!\!^{\prime\prime}$}}

\lefthead{}
\righthead{}
\begin{document}

\title{An Extremely Lithium-Rich Bright Red Giant in the 
Globular Cluster M3\altaffilmark{1}}

\author{
Robert P. Kraft\altaffilmark{2}, 
Ruth C. Peterson\altaffilmark{2}, 
Puragra Guhathakurta\altaffilmark{2,3},
Christopher Sneden\altaffilmark{4}, 
Jon~P.~Fulbright\altaffilmark{2}, 
G. Edward Langer\altaffilmark{5,6}
}

\altaffiltext{1}{Based on observations obtained with the Keck~I 
Telescope of the W. M. Keck Observatory, which is operated by the 
California Association for Research in Astronomy (CARA), Inc.\ on behalf 
of the University of California and the California Institute of Technology.}

\altaffiltext{2}{UCO/Lick Observatory, Dept of Astronomy, University of 
California, Santa Cruz, CA 95064; kraft@ucolick.org, peterson@ucolick.org,
raja@ucolick.org, jfulb@ucolick.org.}

\altaffiltext{3}{Alfred P.\ Sloan Research Fellow}

\altaffiltext{4}{Department of Astronomy and McDonald Observatory, 
University of Texas, Austin, TX 78712; chris@verdi.as.utexas.edu.}

\altaffiltext{5}{Physics Dept, Colorado College, Colorado Springs, 
CO 80903}

\altaffiltext{6}{Deceased 1999 February 16.}

\vskip .5truein
\begin{center}
Accepted for publication in {\it The Astrophysical Journal Letters}
\end{center}

\begin{abstract}
 
We have serendipitously discovered an extremely lithium-rich star on 
the red giant branch of the globular cluster M3 (NGC 5272).
An echelle spectrum obtained with the Keck~I HIRES reveals a \nion{Li}{i}
\wave{6707} resonance doublet of 
520~m\AA\ equivalent width, and our analysis places the star among the 
most Li-rich giants known: log~$\epsilon$(Li)~$\simeq$~+3.0.
We determine the elemental abundances of this star, IV-101, and three 
other cluster members of similar luminosity and color, and conclude that 
IV-101 has abundance ratios typical of giants in M3 and M13 that have 
undergone significant mixing. 
We discuss mechanisms by which a low-mass star may be so enriched 
in Li, focusing on the mixing of material processed by the 
hydrogen-burning shell just below the convective envelope. 
While such enrichment could conceivably only happen rarely, it may 
in fact regularly occur during giant-branch evolution but be 
rarely detected because of rapid subsequent Li depletion.  

\end{abstract}

 
\section{Introduction}

Low-mass giant stars with enhanced lithium are rare. 
Indeed, Li is destroyed in all but the outermost 1--2\% of a 
main-sequence star such as the Sun, and further reduced as the convection 
zone deepens during giant branch evolution (Pilachowski \etal\ 1993).
Yet Li-enhanced giants do indeed exist. 
Wallerstein \& Sneden (1982) discovered that the somewhat metal-rich giant 
HD~112127 has a Li abundance log~$\epsilon$(Li)~= 
+3.0~$\pm$~0.2.\footnote
{[A/B]~$\equiv$~log$_{\rm 10}$(N$_{\rm A}$/N$_{\rm B}$)$_{\rm star}$~--
log$_{\rm 10}$(N$_{\rm A}$/N$_{\rm B}$)$_{\odot}$, and
log~$\epsilon$(A)~$\equiv$~log$_{\rm 10}$(N$_{\rm A}$/N$_{\rm H}$)~+~12.0.}
Brown \etal\ (1989) noted that this is a factor of 30 higher than expected; 
their survey of 644 field giants revealed only one more of similarly high 
Li abundance, plus eight with moderate Li enhancements. 
These stars are of lower mass and luminosity, and are probably less evolved,
than the ``classical'' Li-rich giants of the Galaxy and the Small Magellanic
Cloud (Plez \etal\ 1993).
The possibly distinct group of Li-rich stars associated with dust 
shells is discussed by de~la~Reza \etal\ (1996, 1997).

This paper reports the serendipitous discovery of another extremely 
Li-rich low-mass giant: IV-101, an otherwise normal member of the 
globular cluster M3 (NGC 5272). 
The light-element abundances of IV-101 indicate a moderate 
degree of mixing and its color favors an evolutionary status on 
the first ascent of the giant branch prior to the He core flash.
We suggest that it represents a fleeting phase of Li enrichment 
immediately following a significant mixing event, and that such episodes of 
enrichment are not uncommon among metal-poor and other low-mass giants ---
they are merely short-lived.

\section{Observations and Data Reduction}

The spectroscopic data discussed here are part of an abundance survey 
of 47 M3 giant stars conducted in 1998 March with the Keck~I High 
Resolution Echelle Spectrograph (HIRES --- Vogt \etal\ 1994).
The target stars are largely situated near the center of M3, 
selected from the $VI$ photometry and coordinates
obtained with the post-repair Hubble Space Telescope by Marconi \etal\ 
(1998) and communicated by Ferraro \& Dorman (1998). 
Also included were giants from the outer parts of M3, with photometry from 
Johnson \& Sandage (1956), to provide a nearly complete sample of 
cluster members within 1.5~mag of the red giant branch (RGB) tip. 

With a slit width of $0\farcs86$, the Keck~I HIRES 
achieved a two-pixel spectral resolution of R~=~45,000 
on the Tektronics 2048$\times$2048 detector.  
The S/N ratio was typically 100 per pixel for V~$<$~14. 
Spectral coverage extended from \wave{4470} to \wave{6720} with gaps at the 
ends of each echelle order.  
While the strong \nion{Li}{i} resonance doublet at \wave{6707} was included, 
the weaker \nion{Li}{i} \wave{6104} feature fell in a gap and was not recorded.
We also gathered exposures of quartz and Th-Ar lamps for spectrum 
calibrations, and of rapidly rotating B-type stars for telluric line removal.
Spectrum reduction procedures were identical to those described in
earlier papers of this series (\eg, Sneden \etal\ 1997; Kraft \etal\ 1998).
The equivalent width (EWs) measures are available by request to us and
will be published in the future as part of our larger M3 study.

\begin{figure}[p]
\plotfiddle{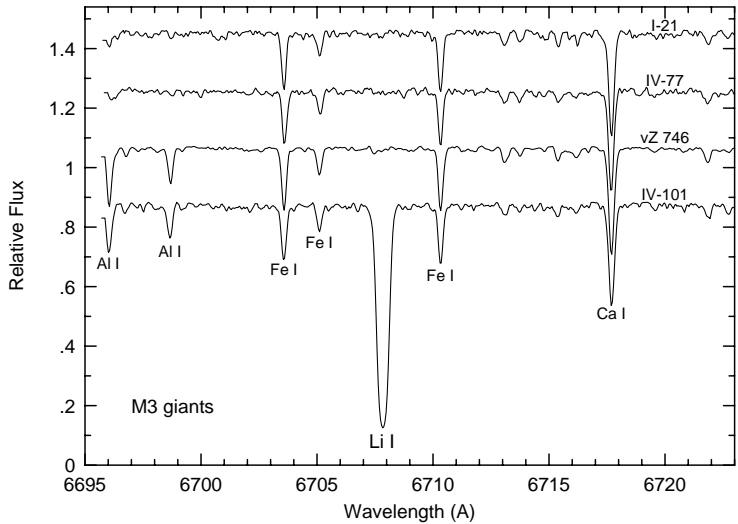}{120pt}{90}{40}{40}{160pt}{-75pt}
\vskip 0.65truein
\caption{Spectra of the \nion{Li}{i} transition region in the
four program stars.
The \nion{Li}{i} EW is $\simeq520$~m\AA\ in IV-101, but the line is
undetected in the other three stars.}
\label{liobs}
\end{figure}

During inspection of the data we spotted an extremely strong 
\wave{6707} \nion{Li}{i} feature in IV-101.
Here we report abundance analyses for IV-101 and three very 
similar M3 giant stars.
Basic data are gathered in Table~1.
All four stars are radial-velocity members of the cluster 
(Gunn \& Griffin 1979; Pryor \etal\ 1988). 
These and our own radial velocity measurements given in Table~1 show no 
velocity variations at the 0.5~\kmsec\ level over periods of decades.
Our spectra in the vicinity of the \nion{Li}{i} features are 
displayed in Figure~\ref{liobs}.
The great strength of the Li feature in IV-101 is obvious 
(EW~$\simeq$~520~m\AA), and the \nion{Al}{i} doublet 
($\lambda\lambda$6696, 6699~\AA) is much stronger in IV-101 than in 
I-21 and IV-77. 
The \nion{Al}{i} line strengths of vZ~746 are comparable to those of IV-101,
yet vZ~746 has no detectable \nion{Li}{i} line.

\section{Abundance Analysis}

\begin{figure}[p]
\plotone{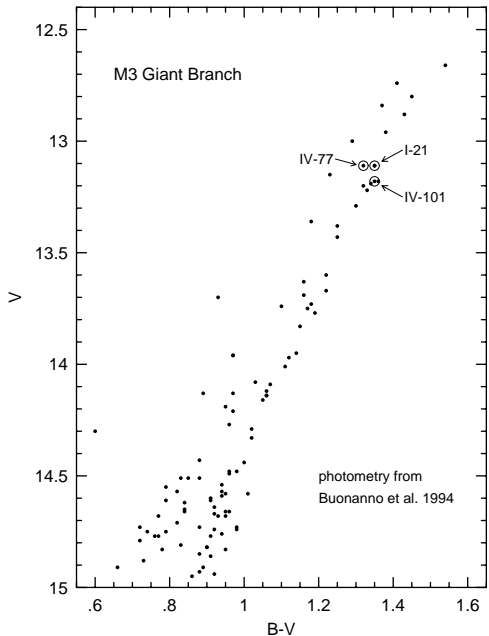}
\caption{The $V$, $B-V$ diagram for the most luminous stars in M3 from the
observations of Buonanno \etal\ (1994).  Three of the program stars are
indicated in the figure, while the fourth star, vZ~746, was not observed by
Buonanno \etal}
\label{buonanno}
\end{figure}

Abundances were derived in the same manner as in
our previous papers (\eg, Sneden \etal\ 1997; Kraft \etal\ 1998).
We first considered the photometric information of the stars.
The positions of IV-101, I-21, and IV-77 (Johnson \& Sandage 1956)
in the M3 CMD of Buonanno \etal\ (1994) are shown in Figure~\ref{buonanno}; 
these stars lie hard against the right-hand edge of the 
RGB--AGB sequence in this figure. 
The CMD and an assumed distance modulus of $(m-M)_0$~=~15.02 
(Djorgovski 1993) yield $M_{\rm bol}$~=~--2.8 for IV-101. 
The central object vZ~746 (von Zeipel 1908) was not observed by 
Buonanno \etal. 
We have estimated its $B-V$ value from its observed $V-I$ and the 
$B-V$ vs $V-I$ relation for other M3 giants (\eg, Marconi \etal\ 1998),
but not plotted it in Figure~\ref{buonanno}.  
vZ~746 is $\sim$0.2~mag brighter than this sequence in the Marconi \etal\ 
$V$ vs.\ $V-I$ diagram.
Probably IV-101, I-21, and IV-77 are on the first-ascent RGB,
but a double-shell AGB phase cannot be entirely ruled out, 
especially for vZ~746.  
None of the four stars was seen to vary in 90 B-band plates obtained
over a two-year period by Corwin \& Carney (1999).

Trial values of \teff\ and log~g were estimated for each star based on 
the earlier calibration (Sneden \etal\ 1992; Kraft \etal\ 1995) of these 
quantities as functions of \MV0\ and \BmV0\ for giants having 
metallicities [Fe/H]~$\sim$~--1.5.  
Final model parameters were determined by iteration, to satisfy 
standard requirements that abundances deduced from iron lines show no 
dependence on line excitation, ionization, or EW.
These parameters are listed in Table~\ref{abunds}.  
All four stars are virtually identical in \teff, log~g, 
v$_{\rm t}$, and [Fe/H]. 
The mean spectroscopic gravity of the four stars, $<$log~g$>$~$\sim$~+0.8, 
agrees well with the mean evolutionary gravity derived from knowledge of the 
M3 distance modulus, an assumed mass of 0.85~\msun, and our values of \teff.

For most elements, the abundances given in Table~\ref{abunds} were derived 
from the EWs, while for Li, N, O, Na, and Ba we used spectral syntheses. 
The \nion{Li}{i} line has doublet structure, and since the line components 
have unequal transition probabilities, an asymmetrical profile is computed.
Yet IV-101's observed line is remarkably symmetrical, possibly indicating 
a small velocity gradient desaturating this very strong line.
Thus our estimate of log~$\epsilon$(Li)~$\simeq$~+3.0 could be an
overestimate.
We found no evidence for $^{\rm 6}$Li, but the anomalous \nion{Li}{i} 
profile precludes a definitive statement on this.
For the other three giants, the estimated upper limit 
log~$\epsilon$(Li)~$\lesssim$~--0.5 is compatible with the low values of 
halo field giants (Pilachowski \etal\ 1993).

For three stars, C abundances were taken, with minor extrapolation, from the 
study of the \wave{4300} CH band in M3 and M13 giants by Smith \etal\ (1996).
For vZ~746, we assumed that 
[C/Fe]$_{\rm vZ~746}$~$\simeq$~[C/Fe]$_{\rm IV-101}$, since these two 
stars have comparable abundances of the other light ``proton-capture''
elements O, Na, Mg and Al.
Abundances of N were then estimated from synthesis of several very 
weak CN lines surrounding the [\nion{O}{i}] lines; these agree well with 
[N/Fe] values based on the calibration of the CN bands using the index 
S(3839) by Smith \etal\ (1996).  
Our derived [N/Fe] in IV-101 is 0.4--0.5~dex larger than those
of the other three stars, but this agrees with Suntzeff's (1981) 
claim that IV-101 has a relatively large N abundances.

Abundances of the remaining elements are comparable in all four stars, and 
very similar to their abundances in other cluster and halo field giants.
The abundances of Eu and Ba are also ``normal'' for the metallicity of 
M3 (Armosky \etal\ 1994; Shetrone 1996a,b). 
The [Ba/Eu]-ratios range from about $-0.2$ to $-0.4$, and the four stars 
do not differ significantly from one another in this ratio. 
Thus there is no evidence for any s-process abundance enhancements.
Overall, the general abundance patterns simply suggest that proton-capture
nucleosynthesis has gone further in IV-101 and vZ~746 than in I-21 
and IV-77: C, O and possibly Mg are depleted, while N, Na, and Al 
are enhanced in IV-101 and vZ~746 compared to I-21 and IV-77. 
These four stars display Na vs.\ O and Al vs.\ Mg anticorrelations, 
symptoms of deep mixing among M3 and M13 giants ({\it cf.}, 
Kraft \etal\ 1997; Shetrone 1996a,b).
Except for the overabundance of Li, IV-101 has essentially the same 
abundances as do well-mixed giants in M3 and M13. 
There are more of these stars in M13 than in M3, but they are not 
rare in either cluster. 

\section{Lithium Production and Depletion in Low-Mass Stars}

It is improbable that M3 IV-101 has retained its primordial Li abundance 
while still occupying a normal place in the M3 color-magnitude array. 
The Li abundance of M3 IV-101 is higher than its probable initial value
of log~$\epsilon$(Li)~$\simeq$~+2.2, that of halo turnoff stars 
(\eg, Spite \& Spite 1982; King \etal\ 1996).
Thus accretion of a substellar body that has not depleted its Li
(e.g.\ Brown \etal\ 1989) cannot provide enough additional Li.
Accretion of Li from a circumstellar disk also appears unlikely.
The presence of substantial cool material would also be seen in the 
\nion{Na}{i} D~lines, whose line formation conditions are similar to those 
of the \nion{Li}{i} 6707~\AA\ doublet. 
But the D~lines in IV-101 are indistinguishable from those of two of 
the other three giants.  

Instead, fresh Li must have been produced and convected into the 
envelope of IV-101.  
This probably happened in a nucleosynthetic ``event'' that included the 
``beryllium transport process'' first proposed by Cameron \& Fowler (1971).
This fusion sequence is
$^{\rm 3}$He($\alpha$,$\gamma$)$^{\rm 7}$Be(e,$\nu$)$^{\rm 7}$Li, with
the requirement that the $^{\rm 7}$Be created in the nuclear processing
zone be transported to cooler envelope regions so that the eventual
$^{\rm 7}$Li product will not be destroyed by proton captures.
Cameron \& Fowler originally wanted this mechanism to operate in 
He-shell fusion zones, in order to simultaneously create the 
Li and s-process overabundances that are observed 
(\eg, Plez \etal\ 1993) in some S and C stars.
But the greatly altered proton-capture element abundances, unaccompanied 
by a barium abundance enhancement, probably rules out a He-fusion
zone origin for the large Li abundance in IV-101.

Li synthesis in a hydrogen shell fusion zone seems more likely.
Deep envelope mixing in low mass stars is explored by 
Sackmann \& Boothroyd (1999), who call their mechanism 
``cool bottom processing'', and focus on its application to the 
Li-rich giants of de~la~Reza \etal\ (1996, 1997). 
They studied both single ``nucleosynthetic event'' and ``steady state'' 
models of various mass/metallicity combinations. 
They argue that the $^{\rm 7}$Li production process, which 
depletes $^{\rm 3}$He, should be even more efficient in globular cluster 
giants than in the giants discussed by de la Reza \etal, since the lower 
metallicity cluster stars have hotter CNO H-burning shells. 

Our abundance results cannot shed light on the physics of the 
nucleosynthesis event (a hydrogen shell-source instability? 
see von~Rudloff \etal\ 1988).
But we can argue that the Li seen in IV-101 was most probably 
synthesized recently, because the Li so produced is likely to be 
destroyed on relatively short time scales. 
In classical models, an M3 star with M$_{\rm bol}$~=~--2.8 has 
T~=~1.1$\times$10$^{\rm 6}$~K at the base of its convective envelope 
(Sweigart \& Gross 1978). 
Li is destroyed by proton capture at temperatures roughly twice this 
value, so even modest overshoot of moving fluid cells below the 
convection zone would bring about Li depletion. 
Such mixing probably occurs quasi-continuously for all stars on the RGB, 
since virtually all bright globular-cluster 
giants show some evidence for enhanced proton-capture products.
To date, Li has been detected in only two other stars out of $\sim$100 
globular cluster giants examined at high spectral resolution.
Thus the depletion time scale for Li in globular-cluster giants 
must be $\lesssim$4$\times$10$^{\rm 4}$~yr, 2\% of the 
evolutionary time scale to the RGB tip of 2$\times$10$^{\rm 6}$~yr 
(Rood 1972; Kraft \etal\ 1993, Fig.~8).  

The two other known Li-rich globular cluster giants differ
somewhat from the abundance pattern of IV-101.
The M5 W~Vir variable V42 (Carney \etal\ 1998) shares the 
low O, high Na, and normal Ba/Eu of IV-101, but neither its Mg nor 
Al are significantly altered, and the Li enhancement is
far less: log~$\epsilon$(Li)~$\simeq$~+1.8.
Li in this AGB star surely must have been produced on the AGB itself, 
again possibly due to H-shell burning, but shortly before the 
star ``looped'' into the W~Vir domain. 
Too much time would have elapsed since the first RGB ascent for Li
synthesized then to have survived in V42.
In NGC~362, Smith \etal\ (1999) discovered a strong 
\nion{Li}{i} line in the giant-branch tip variable V2. 
Their derived abundance is log~$\epsilon$(Li)~$\simeq$~+1.2; other
light elements show no obvious abundance anomalies.
The Sackmann \& Boothroyd (1999) ``cool bottom processing'' may have
been at work in these two stars, but their cases are not as easy to
make as in M3 IV-101.

de~la~Reza \etal\ (1996, 1997) argue that their Li-rich low-mass, 
high metallicity field giants have recently ejected Li-rich dusty shells.
IRAS surveys (\eg, Gregorio-Hetem \etal\ 1993; 
de la Reza \etal\ 1996) detect shells in a large fraction of these stars. 
We looked for evidence for a dusty shell surrounding IV-101 by comparing 
IRAS 100\/\micron\ fluxes centered on each of the four stars listed 
in Table~\ref{abunds}.  
Using the DIRBE-calibrated IRAS maps of Schlegel \etal\ (1998),
we found that the 100\/$\mu$m fluxes at the positions of IV-101, I-21, 
vZ~746, and IV-77 agree to within 15\%.
Unfortunately, 100\/\micron\ flux might not be detectable from a 
shell in which the metallicity is reduced forty-fold. 
We also examined the radial velocity of the deep absorption cores of the 
Na~D lines which, in cool giants, often show evidence of outflow in the 
form of a cold expanding shell (Peterson 1981). 
Relative to neighboring weak \nion{Fe}{i} and \nion{Ti}{i} lines, the core 
displacements are --2.3, --2.4, +2.1, and --2.5~\kmsec, respectively, 
for these four stars, again suggesting that IV-101 is in no way unusual.

We cannot definitely say whether the proposed mixing event is a rare 
one due to individual circumstances, or whether such events commonly occur 
during the RGB and AGB evolution of low-mass giants.
A mixing event that produces a Li enhancement may truly be
an anomaly in globular-cluster stellar evolution. 
The similarity of the CNO and NaMgAl abundance ratios in IV-101 (with 
high Li) and vZ~746 (with undetectable Li) might suggest this. 
Perhaps circulation currents move faster to transport $^{\rm 7}$Be in 
IV-101 than in vZ~746, indicative of a general spread of internal angular 
momenta among cluster giants ({\it cf.} Pinsonneault 1997). 
Since Peterson (1983,1985) showed that BHB stars in M13 have larger
rotational velocities than do both M3 and M5 BHB stars, giants in the 
globular cluster M13 might also be expected to show Li enhancements. 
However, our inspection of the spectra of Kraft \etal\ (1997) revealed 
no prominent \nion{Li}{i} lines in their 18 giants.

Mixing events may be relatively common features of the normal stellar 
evolution of globular-cluster and other low-mass giants, but might be
rarely observed because of the rapid destruction of newly minted Li. 
In this scenario, IV-101 has recently experienced such an event; 
a similar event may also have occurred in vZ~746 but sufficiently far in the 
past ($>$10$^{\rm 4}$~yr) that the Li has by now been completely destroyed. 
Such events probably occur at various luminosities along the giant branch. 
This is consistent with the appearance all along the giant branch 
of changes in CNO and NaMgAl abundance ratios.
Should Li enrichment be a common occurrence, then roughly 1--2\% 
of all such giants should show some Li enhancement. 
Should the He flash be involved, then the red horizontal branch 
stars should show enhancements more frequently. 

\medskip
\bigskip
\bigskip
We thank Franz Ferraro and Ben Dorman for providing new M3 
HST photometry and astrometry, and appreciate the helpful comments
on this work by Inese Ivans and Bruce Carney.
This research was supported by NSF grants AST-9217970 to RPK and JPF, 
AST-9618502 to RCP, and AST-9618364 to CS.
This paper is dedicated to the memory of our beloved colleague Ed Langer,
who died after a brief illness on February 16, 1999.
Ed brought a unique theoretical perspective to our globular cluster
abundance studies.
His career truly embodied the academic ideals of inspiration in both teaching
and research.
He made friends wherever he traveled, and was an inspiration to students.
We will miss him greatly.

\onecolumn
\begin{deluxetable}{lrrrr}
\tablenum{1}
\tablewidth{4.7in}
\tablecaption{M3 Data}\label{abunds}
\tablecolumns{5}
\tablehead{
\colhead{Quantity}                         &
\colhead{vZ 265}                           &
\colhead{vZ 746}                           &
\colhead{vZ 334}                           &
\colhead{vZ 1392}                          \nl
\colhead{}                                 &
\colhead{IV--101}                          &
\colhead{\nodata}                          &
\colhead{IV--77}                           &
\colhead{I--21}                                
}
\startdata
\cutinhead{Basic Data and Model Atmosphere Parameters}
V              &    13.18&      12.96&      13.11&      13.11\nl
B--V           &     1.35&     [1.32]&       1.32&       1.35\nl
T$_{\rm eff}$  &     4200&       4200&       4250&       4200\nl
log g          &     0.85&       0.85&       0.75&       0.75\nl
v$_{\rm t}$    &      1.7&        1.7&        1.6&        1.6\nl
[Fe/H]         &   --1.50&     --1.48&     --1.52&     --1.47\nl
\cutinhead{Absolute log $\epsilon$ Lithium Abundances}
Li (Li I)      &    3.0: & $\leq$--0.5& $\leq$--0.5& $\leq$--0.5\nl
\cutinhead{Relative [X/Fe] Abundances}
C (CH)         &   --0.55&     --0.50&     --0.25&     --0.20\nl
N (CN)         &     +1.0&       +0.6&       +0.5&       +0.5\nl
O ([O I])      &   --0.03&     --0.15&      +0.25&      +0.30\nl
Na (Na I)      &    +0.18&      +0.28&     --0.28&     --0.18\nl
Mg (Mg I)      &    +0.20&      +0.16&      +0.36&      +0.30\nl
Al (Al I)      &    +0.91&      +0.91&     --0.01&     --0.09\nl
Si (Si I)      &    +0.27&      +0.16&      +0.08&      +0.19\nl
Ca (Ca I)      &    +0.24&      +0.27&      +0.29&      +0.27\nl
Sc (Sc II)     &   --0.32&     --0.16&     --0.36&     --0.38\nl
Ti (Ti I)      &    +0.41&      +0.26&      +0.13&      +0.27\nl
V (V I)        &    +0.03&      +0.08&     --0.02&      +0.01\nl
Fe (Fe I)      &   --0.02&      +0.01&      +0.01&      +0.03\nl
Fe (Fe II)     &    +0.02&      +0.00&     --0.01&     --0.02\nl
Ni (Ni I)      &   --0.05&     --0.07&     --0.06&     --0.18\nl
Ba (Ba II)     &    +0.22&      +0.24&      +0.21&      +0.16\nl
Eu (Eu II)     &    +0.63&      +0.48&      +0.39&      +0.44\nl
\enddata
\end{deluxetable}
\twocolumn

\end{document}